# Out-of-body Localization of Virtual Vibration Sources Using a Limited Numbers of Transducers on the Torso

Gen OHARA1, Masashi KONYO1, Satoshi TADOKORO1

1) Tohoku University, Graduate School of Information Sciences (6-6-01 Aramaki Aza Aoba, Aoba-ku, Sendai-shi, Miyagi, Japan 980-8579, konyo@rm.is.tohoku.ac.jp)

Abstract: Stereohaptic vibration is an innovative vibrotactile technology that extends the conventional tactile localization to the surrounding space, representing a virtual vibration source in the external environment. Previously, we have developed displays on the forearms and soles. Today, we present a demonstration of a new jacket-type device, which enables localization at any position around the body by arranging multiple vibrators along the torso centered on the midline. In our demonstration, you can experience the footsteps and roars of a dinosaur walking around you, and it provides an experience that is as if you are in a fantasy movie.

KeyWords： Wearable Device, Vibrotactile Display, Vibrotactile Localization, Phantom Sensation

## 1. Abstract

Distributed vibrations perceived at the body surface may give a sense of the movement and location of vibration sources outside the body. For example, humans perceive the presence of vibration sources in specific directions in daily life, such as sound pressure from music or fireworks. Expressing external objects using vibrotactile localization technology is expected to enhance the sense of presence of objects and increase immersion in virtual and augmented reality.

Conventionally, phantom sensation (PhS) has been used to represent a virtual vibration source on a body surface using a limited number of oscillators. PhS is a tactile illusion phenomenon where a virtual vibration source is perceived between multiple actuated oscillators. Israr et al. propose a method of expressing the sensation of a virtual object moving over the back by having PhS, created by a combination of oscillators arranged in a matrix on the back of a chair, work like virtual oscillators and presenting an apparent motion by the phase difference of vibrations[1]. Tawa et al. have also attempted to extend PhS, which had been limited to the body surface, to the outside of the body[2].

We have been working to realize a "stereohaptic vibration" that enables the localization of virtual vibration sources in three dimensions and the representation of the tactile sensation of an arbitrary object by multiple vibrotactile stimuli on the body surface[3]. Previously, we have developed a bracelet-type display to localize around the forearm and a foot display to represent the two-dimensional localization on the floor. Today, we present a new jacket-type stereohaptic vibration device that enables us to localize virtual vibration sources at any position around the body by arranging multiple vibrators along the torso centered on the midline.

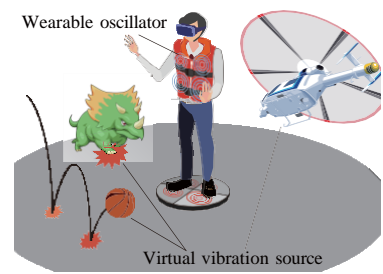

**Fig.1: Conceptual Image of Stereohaptic Vibration**

## 2. Stereohaptic Vibration on the Torso

### 2.1 Principle of Stereohaptic Vibration

We Explain the principle of stereohaptic vibration. In stereohaptic vibration, we quantify the perceived vibration of the virtual source according to the perceptual characteristics of the human vibrotactile senses. The position of the virtual vibration source is localized by distributing the perceived intensity to multiple vibrators, which is like a tactile version of stereophonic sound, an auditory localization method. This section explains the perceived amount of vibrotactile sensations and the method of distributing vibration intensity.

In this study, we used the perceived intensity to quantify the perceived amount of vibrotactile sensation because this parameter considers the frequency dependency of perception of high-frequency vibration in the human pachyderm. We utilized the Intensity Segment Modulation[4], which converts waveforms of arbitrary carrier frequencies while maintaining the tactile sensation of high-frequency vibrations above 100 Hz, considering the low-frequency fluctuation detection characteristics of perceived intensity. During this waveform conversion process, we calculate the perceived intensity of the input waveform.

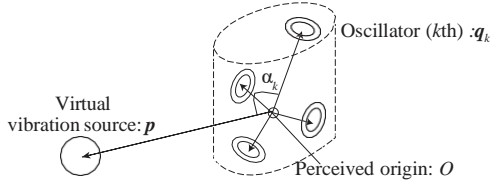

**Fig.2: Schematic image of stereohaptic vibration**

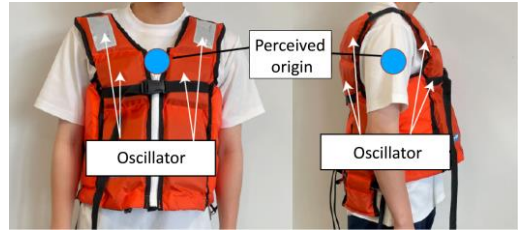

**Fig.3: Stereohaptic vibration jacket**

In stereohaptic vibration, we distribute the perceived intensity to each vibrator, and the vibrators are driven based on the distributed strength to represent the position. Fig.2 shows the positional relationship between the assumed vibrators and the virtual vibration source. In 3D vibration, we represent the source location by distributing the perceived intensity to express the distance and direction. To express the sense of distance, we apply attenuation to the source vibration according to the distance $|\overline{p}|$ between the virtual vibration source $p$ and the perception origin $O$. When the vibration source generates vibrations with a perceived intensity $I_0$, the perceived intensity $I$ at the perception origin position is obtained using the distance attenuation function $d$ as shown in Equation (1). Here, the perception origin $O$ is defined as the central position of the vibration image obtained by PhS when $N$ vibrators attached to the body are driven equally.

$$I = d(\overline{p}) \cdot I_0 \quad (1)$$

To express the sense of direction, the intensity $I$ perceived at the perceived origin is distributed based on the positional relationship between the oscillator and the vibration source. The distribution ratio $r_k$ of perceived intensity for the $k$-th oscillator is determined as in equation (2). The sum of the distribution ratio $r_k$ is kept at 1 to maintain the distance sensation.

$$I_k = r_k \cdot I \quad (2)$$

The $r_k$ is determined from the inner product $u_k$ (equation (3)) of the source vector $p$ and the oscillator vector $q_k$ centered at the perceived origin $O$. This expresses, for example, that the vibration is strongest when the direction of the vibration source and the oscillator coincides with the direction of the perceived origin and is weakest in the opposite direction.

$$u_k = \overline{q}_k \cdot \overline{p} = \frac{q_k \cdot p}{\|q_k\| \|p\|} \quad (3)$$

### 2.2 Vibration Jacket

In this study, we constructed a stereohaptic vibration system using a jacket-type device (Fig.3) that expresses a sense of localization around the human body.

This device takes the form of a jacket embedded with a total of eight oscillators(ACTUATOR639897 Foster electric), four on the front and four on the back. In this setup, the oscillators are arranged so that the perceived origin is on the midline, as shown in Figure 3. Adjusting the belt makes it possible to distribute the force evenly with which the oscillators are pressed against the body.

Controlling the vibrations based on the principles in the previous section, the virtual vibration source is localized at any position in surrounding space.

### 3. System Configuration

We explain the system configuration (Fig.4). To represent vibrations, amplitude-modulated waves were generated from audio signals generated in Unity based on the stereohaptic vibration method, converted to analog signals using a USB audio interface (MOTU, Ultralite AVB), amplified by an amplifier, and used to drive eight actuators in the vibration jacket.

The participants wear a head-mounted display (VIVE Pro eye) to see the VR vision and hear stereophonic sound output from Unity.

The current stereohaptic vibration system requires some wired connection devices, which limits its convenience and portability. We are planning to implement wireless control for the oscillators so that we can enhance portability and enable the development of a more expandable system in the future.

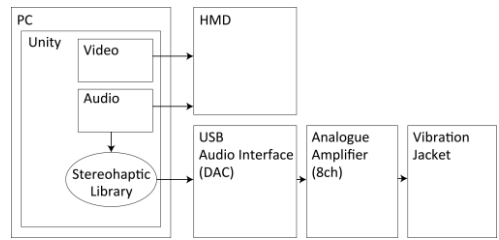

**Fig.4: System configuration**

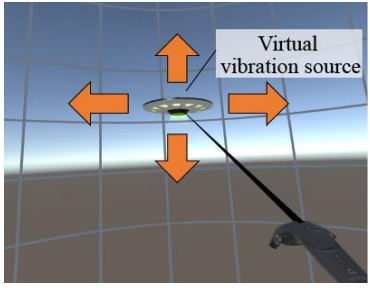

**Fig.5: Demonstration overview: Participants can feel the motion of the virtual vibration source moving around the body**

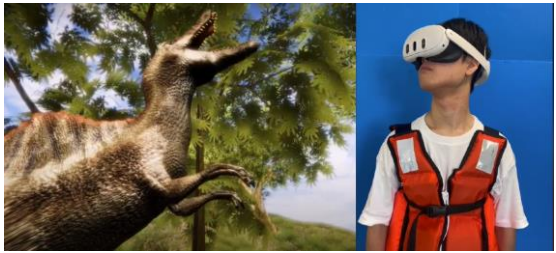

**Fig.6: Combining the stereohaptic vibration with VR contents**

## 4. Demonstration

In our demonstration, we provide a passive and active vibrotactile localization experience.

Participants can move the virtual vibration sources to arbitrary positions and localize (Fig.5). The sound effects and vibrations change according to the type of object, allowing participants to experience the presence of various virtual objects.

We also have combined the stereohaptics with VR content (Fig.6). Simulating the vibrations of a dinosaur's footsteps and roar as it moves in virtual reality, participants can feel as if they have entered the world of a movie or game.

## 5. Conclusion

We have confirmed that our method can present three-dimensional stereohaptic vibration and enhance the realism of virtual reality. In our demonstration, you can experience the presence of the virtual vibration objects and immersive virtual reality.

## 6. Acknowledgement

This work was supported in part by JSPS KAKENHI Grant Number JP21H05795.